\documentclass[preprints,article,accept,pdftex,moreauthors]{Definitions/mdpi} 
\firstpage{1} 
\pubvolume{1}
\issuenum{1}
\articlenumber{0}
\pubyear{2026}
\copyrightyear{2026}
\datereceived{ } 
\daterevised{ } 
\dateaccepted{ } 
\datepublished{ } 
\pdfoutput=1 

\Title{Generalized Beth-Uhlenbeck Approach to the 2+1D Gross-Neveu Model}

\Author{Biplab Mahato $^{1}$\orcidA{} and David Blaschke $^{1,2,3}$\orcidB{}*}

\AuthorNames{Biplab Mahato and David Blaschke}

\address{%
$^{1}$ \quad Institute of Theoretical Physics, University of Wrocław, Plac Maxa Borna 9, 50-204 Wrocław, Poland\\
$^{2}$ \quad Center for Advanced Systems Understanding (CASUS), Untermarkt 20, D-02826 Gorlitz, Germany \\
$^{3}$ \quad Helmholtz-Zentrum Dresden-Rossendorf (HDZR), Bautzener Landstrasse 400, D-01328 Dresden, Germany}

\corres{Correspondence: biplab.mahato@uwr.edu.pl}

\abstract{We study the thermodynamics of the (2+1) dimensional Gross-Neveu model inspired from graphene. We focus on the entropy density of the Gaussian fluctuation beyond the mean field. The full in-medium, momentum-dependent evaluation reveals that the fluctuations give a substantial contribution, even comparable to that of the mean field. We argue that the back-reaction from the fluctuations to the mean field should be included, which reduces the contribution mainly coming from the Landau-damping region. To treat this self-consistently, we use the generalized version of the Beth-Uhlenbeck approach for the entropy density. Compared with the standard Beth-Uhlenbeck formulation, the generalized version suppresses the low-energy contributions while preserving the bound-state effects. The fractional entropy carried by bound excitons and free fermions reveals a sharper crossover of the degrees of freedom in the generalized version, which is consistent with Mott-transition physics in two-dimensional materials.}


\protect 
\begin{document}


\section{Introduction}
The Gross-Neveu (GN) model is a quantum field theory of $N$ Dirac fermion species characterized by a four-fermion contact interaction. 
Originally it was introduced by David Gross and André Neveu in 1974 \cite{Gross:1974jv} in $(1+1)$ dimensions to demonstrate asymptotic freedom and spontaneous mass generation. 
Its $(3+1)$-dimensional counterpart was introduced thirteen years earlier by Yoichiro Nambu and Giovanni Jona-Lasinio \cite{Nambu:1961tp} as a model of locally interacting nucleons to describe the mass gap. Later, as quarks were being established as the fundamental degrees of freedom, the nucleonic degrees of freedom were replaced by the quark ones \cite{Eguchi:1976iz, Kikkawa:1976fe}. 
Since then, both models have been widely studied. 
The NJL model, as a low-energy effective field theory of Quantum Chromo-dynamics (QCD), is employed to model the equation of state at extreme conditions as well as to study the QCD phase diagram. The $1+1$D model, which is renormalizable unlike the NJL model, is used as a playground for investigating various non-perturbative phenomena in interacting fermionic systems. 
The $(2+1)$ dimensional version of the Gross Neveu model serves as a middle ground, as it shows complex phase boundaries \cite{Rosenstein:1988dj} while being renormalizable \cite{Rosenstein:1988pt} within the $1/N$ expansion \cite{tHooft:1973alw}. 
The model attracted renewed interest after the discovery of graphene \cite{Novoselov:2004xxs} and the subsequent rise of two-dimensional Dirac materials. The low-energy excitations near the high-symmetry points in the Brillouin zone (Dirac cone) of graphene behave as massless Dirac particles \cite{Novoselov:2005kj}. The GN model provides a natural framework for describing the generation of an insulating gap in such systems \cite{lopes:2022exc}. In \cite{Ebert:2018dzs}, the authors considered a specific version of the model respecting the symmetries of graphene. The paper presents the mean field approximation and the Gaussian fluctuations for the correlations. The Gaussian fluctuations and their effect on the thermodynamics were cast into the Beth-Uhlenbeck formalism following previous work on the NJL model, like in \cite{Hufner:1994ma}. In our previous work \cite{Mahato:2024fta}, we numerically investigated the model. In particular, we showed the finite momentum effects on the fluctuations, namely, that bound states reappear when their momenta relative to the thermal system exceed the medium-dependent Mott momentum and that Landau damping modes exist at lowest energies. The latter contribute substantially to the thermodynamics, sometimes being comparable to the mean fields. It is worth noting that the back-reaction, i.e., the effect of the fluctuations back to the mean field, is not considered in this framework. A self-consistent treatment of the thermodynamics is therefore required. There are a few examples \cite{Rochev:2009zz, CamaraPereira:2020ipu, Blaschke:2017boi} on how to include the effect of the correlations back to the mean field. Here, we will focus on the $\Phi$-derivable approach to quantum field theory \cite{Baym:1961zz} which contains the generalized version of the Beth-Uhlenbeck one for a special choice of the $\Phi$ functional. We will demonstrate the suppression of the thermodynamic contribution from low-lying Landau damping due to the back-reaction. In this article, we will compare the thermodynamical quantity entropy density for the Beth-Uhlenbeck approach with its generalized version. We also present the entropy fraction of bound excitons and constituent fermions as a measure for the medium dependence of the composition, comparable to the ionization degree discussed for nuclear \cite{Ropke:1983lbc,Blaschke:2023pqd} and electromagnetic \cite{Ebeling:2008mg,Ropke:2018ewt} plasmas.

The article is structured as follows. In the next chapter, we will introduce the model and very briefly present the method of mean field approximation along with Gaussian fluctuations. The chapter after that sketches the shortcomings of this method and makes a case for the generalized Beth-Uhlenbeck approach. The following section provides some numerical results for the model introduced in the second chapter to illustrate the point. Finally, we summarize and discuss the implications and future work.

\section{Model and Methodology}
The Lagrangian for the $(2+1)$D Gross-Neveu model consists of the free Dirac part with a four-fermion interaction of the form $(\bar{\psi}\Gamma_i\psi)^2$, where the vertices $\Gamma_i$ reflect internal symmetries in different channels, typically a combination of Dirac gamma matrices, 

\begin{equation}
    \mathcal{L} = \bar{\psi}(i\gamma^\nu\partial_{\nu} - m_0 -\gamma^0\mu)\psi - \sum_{i=1}^{4}\frac{G}{2N}\left(\bar{\psi }\Gamma_i\psi\right)^2.
\end{equation}
The specific interaction channels that we consider are inspired by graphene. See, for example, the references \cite{Gusynin:2007ix, Ebert:2018dzs, Mahato:2024fta} for further details. The first reference discusses different symmetry channels available to graphene, while the second one chooses the interaction channels $\Gamma_i = \{I, \gamma_{45}, i\gamma_5, i\gamma_4\}$, which are used in our previous work as well as in this article. The coupling constants are taken to be the same constant $G$. It is possible to consider different coupling constants for different channels, which results in the occurrence of non-zero condensates in different channels (different phases) in the vacuum; see \cite{Zhukovsky:2015ncz} for a discussion. With the particular choice of high symmetry in the present work, we have a non-zero condensate at the mean-field level only in the scalar channel.

To obtain the partition function, we first use a Hubbard-Stratonovich transformation by introducing auxiliary fields, $\varPhi_i$.
\begin{equation}\label{eqn:partition_function}
	\mathcal{Z} = \int\prod_{i=1}^4\mathcal{D}\varPhi_i \textrm{exp}\left\{ -N\int d^3x \sum_{i=1}^4\frac{\varPhi_i^2}{2G} + N \textrm{Tr}\ln S^{-1}\right\},
\end{equation}
with  $S^{-1} = \gamma_\mu\partial_\mu - m_0 - \mu\gamma_0 + \sum_{i=1}^4\Gamma_i\varPhi_i$. Now, we split the auxiliary fields into a mean field and a fluctuation part, 
$\varPhi_i \to \bar{\varPhi}_i + \delta\varPhi_i$. 
The mean field is assumed to be constant over space-time, while the fluctuations are considered small compared to the mean-field values. Then, the partition function can be written up to the second order in the expansion of $\delta\varPhi_i$ as $\mathcal{Z} = \mathcal{Z}_{\rm mf}\mathcal{Z}_{\rm fl}$, where the mean-field part is
\begin{equation}
    \mathcal{Z}_{\textrm{mf}} = \textrm{exp}\left\{-N\int d^3 x\sum_{i=1}^4\frac{\bar{\varPhi}_i^2}{2G} - N\textrm{Tr}\ln S_{\textrm{mf}}^{-1}\right\},
\end{equation}
with $S_{\textrm{mf}}^{-1} = S^{-1}\left(\varPhi_i \to \bar{\varPhi}_i\right)$. 
The fluctuation part is
 \begin{equation}\label{eq:partition_fluc}
	\mathcal{Z}_{\textrm{fl}} = \int\prod_{i=1}^4\mathcal{D}\delta\varPhi_i \textrm{exp}\left\{ - \sum_{i=1}^4\frac{\delta\varPhi_i^2}{2}\left(\int d^3x\frac{N }{G} + 2N\textrm{Tr}\ln(1 + S_{\textrm{mf}}\Sigma)\right)\right\},
\end{equation}
where we introduced the notation 
$\Sigma = \sum_{i=1}^4\Gamma_i\delta\varPhi_i$. 
In the mean field approximation, we first ignore the fluctuation term $\mathcal{Z}_{\textrm{fl}}$ and find the 
mean-field value by extremizing the thermodynamic potential $\Omega_{\textrm{mf}}=-({1}/{V})\ln\mathcal{Z}_{\textrm{mf}}$, 
where $V$ is the volume ($V=\beta \ell^2$, $\beta$ is the inverse temperature and $\ell$ is the length scale), 
${\delta \Omega_{\textrm{mf}}}/{\delta \varPhi_i}=0$. 
The result of this procedure is also known as the gap equation.
The fluctuations are calculated with the mean-field values obtained in the previous step. In the paper \cite{Hufner:1994ma} the authors suggest to consider a generalized version of the gap equations where one also includes the fluctuation parts of the thermodynamical potential in the calculation of the gap equation, i.e. $\frac{\partial (\Omega_{\textrm{mf}} + \Omega_{\textrm{fl}})}{\partial \varPhi_i} = 0$.  However, in the present work, we restrict ourselves to the consideration of the former case only. 

The fluctuation term is simplified further by expanding the log term in \eqref{eq:partition_fluc} as $\textrm{Tr}\ln(1+S_{\textrm{mf}}\Sigma) \approx  \textrm{Tr}(S_{\textrm{mf}}\Sigma) -\frac{1}{2} \textrm{Tr}(S_{\textrm{mf}}\Sigma S_{\textrm{mf}}\Sigma)$. 
The linear term vanishes due to the gap equations, and we can evaluate the Gaussian integral to obtain,
\begin{equation}
    \Omega_\textrm{fl} = -\frac{1}{V}\ln\mathcal{Z}_\textrm{fl}= \frac{1}{2}\sum_{i=1}^4\textrm{Tr}\ln\left( \frac{N}{G}  - \Pi_i\right), 
\end{equation}
where we have defined the polarization function $\Pi_i = -\frac{N}{V}\textrm{Tr}(S_{\textrm{mf}}\Gamma_iS_{\textrm{mf}}\Gamma_i)$.

The expression can be cast into the Beth-Uhlenbeck form, see, for example, the case of the NJL model in \cite{Zhuang:1994dw},
\begin{equation}\label{eqn:omega_fl}
    \Omega_{\textrm{fl}} = -\sum_{i}\int \frac{d^2q}{(2\pi)^2} \int_{-\infty}^{\infty} \frac{d\omega}{2\pi} g(\omega) \delta_i(\omega, q),
\end{equation}
where the phase shift is $\delta_i(\omega, q) = \textrm{Im}\ln\left(\frac{N}{G} - \Pi_i(\omega + i\eta, q)\right)$ and  $g(\omega) = (e^{\beta \omega} - 1)^{-1}$ is the Bose distribution function.

The entropy density, which is the temperature derivative of the pressure ($\mathcal{P} = -\Omega$), can be derived to have the form,

\begin{equation}\label{eq:entropy_BU}
    \mathcal{S}_{\textrm{fl, BU}} = \sum_{i}\int \frac{d^2q}{(2\pi)^2} \int_{-\infty}^{\infty} \frac{d\omega}{2\pi} \frac{\partial g(\omega)}{\partial T} \delta_i(\omega, q)~.
\end{equation}

\section{Generalized Beth-Uhlenbeck Approach}
The phase shift defined in the previous section at zero momentum is finite (strictly between $0$ and $\pi$) only for frequencies above the threshold ($2m$). 
If there is a bound state, then the phase shift jumps by $\pi$ between the bound state mass and the threshold \cite{Levinson:1949os}. 
Frequencies below the bound state mass it is strictly zero. However, at a finite momentum, we see the appearance of a non-trivial bump in the space-like region ($\omega < q$) also known as the Landau-damping region; see figure 8(a) in \cite{Mahato:2024fta} for an illustration. 
The small phase shifts in this region pick up disproportionately large contributions to the thermodynamical quantities due to the divergence of the bosonic distribution function at small energies (see equation \eqref{eqn:omega_fl}). 
As noted in our previous work, as well as in the case of the Polyakov loop coupled NJL model in \cite{Maslov:2023boq}, these contributions can be comparable to the mean field values. 
It is worrying if the terms that we obtained by expanding around the mean-field compare or even surpass the mean-field contribution, as it questions the very validity of such an expansion scheme. 

\begin{figure}
    \centering
    \includegraphics[width=0.6\linewidth]{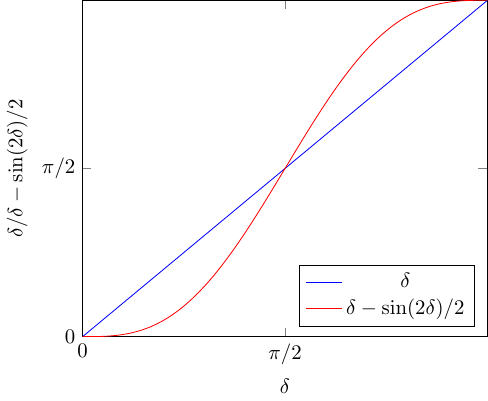}
    \caption{Comparison of the contribution coming from the phase shift in the Beth-Uhlenbeck approach and in the generalized Beth-Uhlenbeck approach. The generalized version suppresses the small phase shift and enhances the phases near $\pi$.}
    \label{fig:phases_gen_phases}
\end{figure}

The significant contributions to thermodynamics come from the lower frequencies of the correlated states. 
It should be noted that the composite states are constructed from fermionic degrees of freedom, and at such low energies, well below the threshold, treating the correlation as a strictly elementary boson might be inadequate and risks overcounting. 
Some parts of the correlations should be treated as corrections to the mean field. 
This is termed the back-reaction, i.e., the effect of the correlation back to the mean field. 
There are several attempts to quantify such effects for the case of the NJL model and its variants. 
For example see \cite{rochev:2009mes, pereira:2020one, Radzhabov:2010dd, kumar:2025int} etc. 
However, in this article, we will focus on the correlations. 
To properly account for the thermodynamic contribution by the correlations, it is necessary to use a self-consistent treatment of the theory. 
One such treatment is the $\Phi$-derivable approach \cite{Baym:1961zz}. 
Within this approach, the thermodynamic potential is constructed such that it remains stationary under variations of the propagators, ensuring conservation laws and consistency. 
It is possible to cast the entropy density in a similar form as in equation \eqref{eq:entropy_BU} \cite{Blaschke:2025eyt},

%
\begin{equation}\label{eq:entropy_genBU}
    \mathcal{S}_{\textrm{fl, gBU}} = \sum_{i}\int \frac{d^2q}{(2\pi)^2} \int_{-\infty}^{\infty} \frac{d\omega}{2\pi} \frac{\partial g(\omega)}{\partial T} \left(\delta_i(\omega, q) - \frac{\sin(2\delta_i(\omega, q))}{2} \right)~.
\end{equation}
The additional term $- \sin(2\delta)/2$ implements the desired correction due to back-reaction. 
For small phase shifts, i.e. the scattering states and the soft Landau damping modes, this correction strongly suppresses their contribution. 
Near the bound state pole, the term vanishes, preserving all the effects of the bound state contributions. We illustrate this in figure \ref{fig:phases_gen_phases}, by plotting the phase shift ($\delta$) and the generalized phase shift ($\delta - \sin(2\delta)/2$).

\section{Results}
For the numerical calculations, we use the model parameters from our previous work in \cite{Mahato:2024fta}. 
The renormalized coupling constant ($g$) is defined as $\frac{1}{g} = \frac{1}{G} - \frac{\Lambda}{\pi}$. 
The mass scale defined by $M = \frac{\pi}{|g|}$ sets the scale of the theory. 
We use this scale for all the figures presented in this article.  
The cutoff used in this article is $\Lambda = 5M$. 
Finally, the free mass parameter $m_0$ is taken such that $\kappa = m_0/G = 0.046 M^2$, which gives the vacuum value of pseudoscalar mass of $0.1M$. 
It is also possible to consider the chiral limit $m_0 = 0$. 
However, it is not presented here. 
The numerical code used in our work is publicly available \cite{Mahato:2025gh} and can be used to obtain the results in the chiral limit. 
Note that in equation \eqref{eq:entropy_BU} the momentum $q$ is the momentum of the collective degrees of freedom and not of the constituent fermions. 
So the cutoff required for the integral to be regulated may differ from the cutoff $\Lambda$ used as the regulator for the fermions. 
We follow in this article the convention from \cite{Maslov:2023boq} and take the cutoff in the range $[\Lambda, 2\Lambda]$.

\begin{figure}
    \centering
    \subfloat[\centering]{\includegraphics[width=0.45\textwidth]{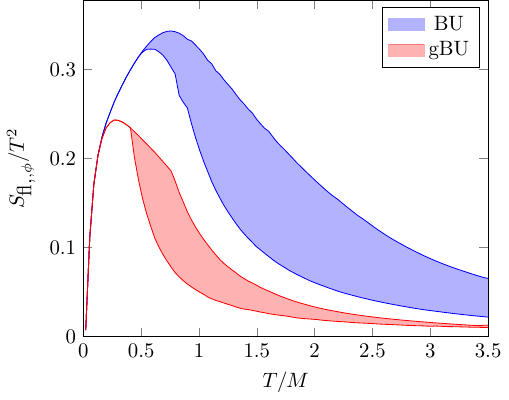}}
    \hfill
    \subfloat[\centering]{\includegraphics[width=0.45\textwidth]{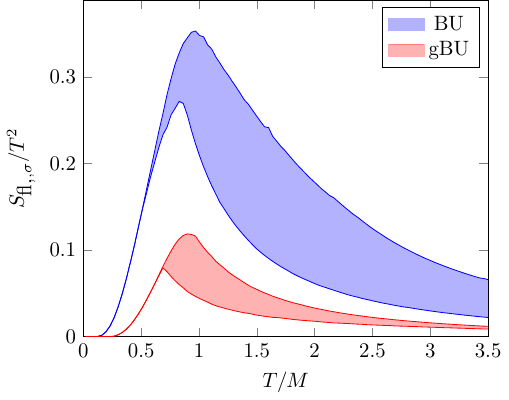}}
    \caption{Entropy density fluctuations for pseudoscalar (\textbf{a}) and scalar (\textbf{b}) channels in the Beth-Uhlenbeck (blue) and the generalized Beth-Uhlenbeck (red) approach.}
    \label{fig:fluctutation_entropy_channels}
\end{figure}

In figure \ref{fig:fluctutation_entropy_channels} we show the entropy density fluctuations in Beth-Uhlenbeck (blue) (using equation \eqref{eq:entropy_BU}) and generalized Beth-Uhlenbeck (red) (using equation \eqref{eq:entropy_genBU}) approaches for scalar (b) and pseudo-scalar channels (a). Both figures show a band for the entropy density, corresponding to the correlation cutoff between $\Lambda$ and $2\Lambda$. 
The lower line of the band represents the lower cutoff, and the upper one represents the larger cutoff. 
We note that the contribution to the fluctuation is significantly higher in the Beth-Uhlenbeck approach. 
We also note that at low temperature, as we have bound states in the pseudo-scalar channel, the generalized Beth-Uhlenbeck formula does not suppress this region. It can be seen from the part where the entropy density in both approaches is very close to each other at small temperatures.

%
\begin{figure}
    \centering
    \includegraphics[width=0.6\linewidth]{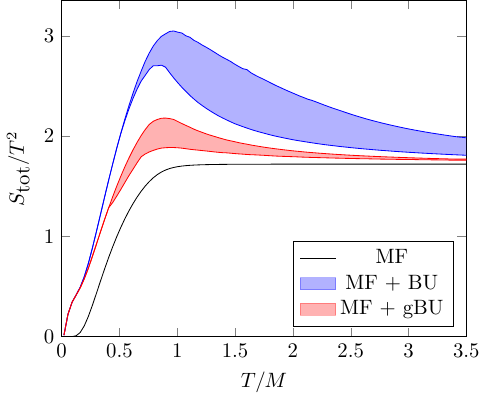}
    \caption{Total normalized (dimensionless) entropy density for the case of Beth-Uhlenbeck (blue) and generalized Beth-Uhlenbeck formalism (red).}
    \label{fig:ent_fl_tot}
\end{figure}
In Figure \ref{fig:ent_fl_tot}, we compare the total entropy fluctuation density for both approaches. 
Once again, we note that the difference is minimal at small temperatures and also in the high temperature limit, where all the fluctuations die out, leaving only the mean-field part. 
Only in the intermediate region of temperatures ($T\sim M$), these two methods produce dissimilar results. 
It becomes clearer if one considers the compositions, i.e., the fraction of entropy carried by each individual degree of freedom. 
The pseudoscalar collective particles dominate at low temperature, while the constituent fermion species have the dominant contribution to the entropy as the collective modes melt at increasing temperatures. 
The scalar modes only have a significant contribution slightly below the Mott temperature. 
It is instructive to compare this result with the ionization degree of strongly correlated fermionic systems undergoing the Mott transition. 
For example, see \cite{asano:2014exc,steinhoff:2017exc, wrona:2022pai} for two-dimensional materials, and \cite{schmidt:1990gen} for the case of nuclear matter. 
The papers show the ionization degree, i.e., the fraction of free electrons in the mixture of free electrons and bound excitons, as a function of density. 
Below and above the Mott transition, the excitons and the free electrons dominate, respectively. 
There is a sharp transition from one to the other at the Mott transition density. 
The generalized Beth-Uhlenbeck approach shows a sharper transition of degrees of freedom. 
It is worth noting that in realistic excitonic systems, the ionization degree remains finite even at very low densities and approaches zero only at a finite density before the Mott transition sets in. 
In contrast, GN and NJL-type models effectively incorporate a confinement-like mechanism. 
As a result, at very low temperatures, only the bound excitons exist as relevant degrees of freedom, which translates to zero ionization degree. 
This qualitative difference reflects the limitations of contact-interaction models in capturing the physics of dilute exciton gases at extremely low densities.

\begin{figure}
    \centering
    \includegraphics[width=0.6\linewidth]{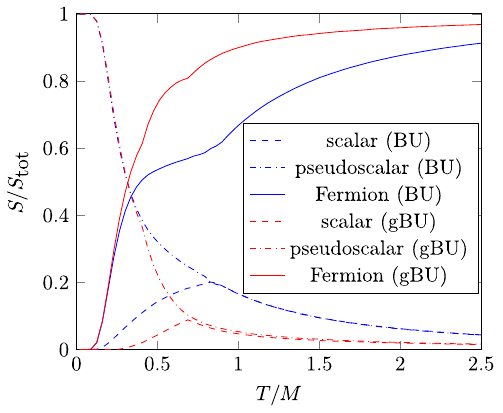}
    \caption{Fraction of the total entropy carried by the scalar, pseudo-scalar channels and the constituent fermions. The figure is shown for the collective mode cutoff of $\Lambda$. For the other cutoffs, the qualitative feature of the figure remains unchanged.}
    \label{fig:composition}
\end{figure}



\section{Discussion}
We have computed and compared the entropy density of the fluctuations in the Beth-Uhlenbeck approach and its generalized version. 
The contribution from the fluctuations (in BU) is significant, which warranted us to include back-reaction and to look for a self-consistent treatment. 
The generalized Beth-Uhlenbeck formula, derived from the $\Phi$-derivable approach, includes the back-reaction of the two-particle correlations onto the mean-field. 
By subtracting the term $\frac{1}{2}\sin(2\delta_i)$, it strongly suppresses the contribution from the continuum states, while preserving the thermodynamics of the bound excitons. 
The result is a significant reduction in the entropy fluctuations at intermediate temperatures $T \sim M$. 
Low and high temperature regimes are largely unaffected. Consequently, it shows a sharper crossover from bound excitons to free fermions, which is a signature of Mott-like dissociation of the bound state to the plasma of constituent fermions. This behavior mirrors the ionization degree observed in  theoretical and experimental studies of excitons in two-dimensional materials. 

The article demonstrates that the generalized Beth-Uhlenbeck formulation offers a thermodynamically more consistent description of strongly correlated fermionic systems. 
Nevertheless, there are a few limitations that should be addressed. The effect of back-reaction on correcting the mean-field needs to be addressed. 
The cutoff dependence for the collective mode integral persists, although it is noticeably weaker in the gBU case. 
Usage of a cutoff scheme other than the 3D sharp cutoff used here, 
or usage of non-local coupling, can remove such dependence. 
Finally, the analysis was carried out at zero chemical potential. The inclusion of finite chemical potentials should shed light on density-driven Mott-dissociation physics due to phase space occupation (Pauli blocking) as well.

\vspace{6pt} 





\authorcontributions{
``Conceptualization, D.B. and B.M.; 
software, B.M.;
writing---original draft preparation, B.M.; writing---review and editing, D.B.; visualization, B.M.; supervision, D.B.. All authors have read and agreed to the published version of the manuscript.''}

\funding{B.M. acknowledges a stipend from the International Max Planck Research School for Quantum Dynamics and Control at the Max-Planck Institute for Physics of Complex Systems in Dresden, Germany. The work of D.B. and B.M. was supported by NCN under grant No. 2021/43/P/ST2/03319.}

\dataavailability{The data used to plot the figures in the paper is produced using the code authored by B.M. which is publicly available at https://github.com/biplab37/PhaseGN.}

\acknowledgments{The authors acknowledge discussions and collaboration with Gerd R\"opke and Dietmar Ebert.}

\conflictsofinterest{The authors declare no conflicts of interest.} 



\abbreviations{Abbreviations}{
The following abbreviations are used in this manuscript:
\\

\noindent 
\begin{tabular}{@{}ll}
GN & Gross-Neveu\\
NJL & Nambu-Jona-Lasinio\\
QCD & Quantum Chromodynamics\\
BU & Beth-Uhlenbeck\\
gBU & generalized Beth-Uhlenbeck\\
NCN & Narodowe Centrum Nauki
\end{tabular}
}

\appendixtitles{no} 



\reftitle{References}


\bibliography{references}

@article{Ropke:1983lbc,
    author = {R{\"o}pke, G. and Schmidt, M. and M{\"u}nchow, L. and Schulz, H.},
    title = "{Particle clustering and Mott transition in nuclear matter at finite temperature (II)}",
    doi = "10.1016/0375-9474(83)90265-8",
    journal = "Nucl. Phys. A",
    volume = "399",
    pages = "587--602",
    year = "1983"
}

@article{Blaschke:2023pqd,
    author = {Blaschke, D. and Cierniak, M. and Ivanytskyi, O. and R{\"o}pke, G.},
    title = "{Thermodynamics of quark matter with multiquark clusters in an effective Beth-Uhlenbeck type approach}",
    eprint = "2308.07950",
    archivePrefix = "arXiv",
    primaryClass = "nucl-th",
    doi = "10.1140/epja/s10050-023-01229-8",
    journal = "Eur. Phys. J. A",
    volume = "60",
    number = "1",
    pages = "14",
    year = "2024"
}

@article{Ebeling:2008mg,
    author = "Ebeling, W. and Blaschke, D. and Redmer, R. and Reinholz, H. and Ropke, G.",
    title = "{The Influence of Pauli blocking effects on the properties of dense hydrogen}",
    eprint = "0810.3336",
    archivePrefix = "arXiv",
    primaryClass = "physics.plasm-ph",
    doi = "10.1088/1751-8113/42/21/214033",
    journal = "J. Phys. A",
    volume = "42",
    pages = "214033",
    year = "2009"
}

@article{Ropke:2018ewt,
    author = {R{\"o}pke, Gerd and Blaschke, David and D{\"o}ppner, Tilo and Lin, Chengliang and Kraeft, Wolf-Dietrich and Redmer, Ronald and Reinholz, Heidi},
    title = "{Ionization potential depression and Pauli blocking in degenerate plasmas at extreme densities}",
    eprint = "1811.12912",
    archivePrefix = "arXiv",
    primaryClass = "physics.plasm-ph",
    doi = "10.1103/PhysRevE.99.033201",
    journal = "Phys. Rev. E",
    volume = "99",
    number = "3",
    pages = "033201",
    year = "2019"
}

@article{tHooft:1973alw,
    author = "'t Hooft, Gerard",
    editor = "Taylor, J. C.",
    title = "{A Planar Diagram Theory for Strong Interactions}",
    reportNumber = "CERN-TH-1786",
    doi = "10.1016/0550-3213(74)90154-0",
    journal = "Nucl. Phys. B",
    volume = "72",
    pages = "461",
    year = "1974"
}

@article{schmidt:1990gen,
  title = {Generalized Beth-Uhlenbeck Approach for Hot Nuclear Matter},
  author = {Schmidt, Martin and R{\"o}pke, Gerd and Schulz, Hartmut},
  year = 1990,
  journal = {Annals of Physics},
  volume = {202},
  number = {1},
  pages = {57--99},
  issn = {0003-4916},
  doi = {10.1016/0003-4916(90)90340-T}
}

@article{Radzhabov:2010dd,
  title = {Nonlocal {{PNJL}} Model beyond Mean Field and the {{QCD}} Phase Transition},
  author = {Radzhabov, A.E. and Blaschke, D. and Buballa, M. and Volkov, M.K.},
  year = 2011,
  journal = {Physical Review D},
  volume = {83},
  number = {11},
  eprint = {1012.0664},
  primaryclass = {hep-ph},
  pages = {116004},
  publisher = {American Physical Society},
  doi = {10.1103/PhysRevD.83.116004},
  archiveprefix = {arXiv}
}

@article{pereira:2020one,
  title = {One-Meson-Loop {{NJL}} Model: {{Effect}} of Collective and Noncollective Excitations on the Quark Condensate at Finite Temperature},
  shorttitle = {One-Meson-Loop {{NJL}} Model},
  author = {Pereira, Renan C{\^a}mara and Costa, Pedro},
  year = 2020,
  journal = {Physical Review D},
  volume = {101},
  number = {5},
  pages = {054025},
  publisher = {American Physical Society},
  doi = {10.1103/PhysRevD.101.054025}
}

@article{rochev:2009mes,
  title = {Meson Contributions in the {{Nambu-Jona-Lasinio}} Model},
  author = {Rochev, V. E.},
  year = 2009,
  journal = {Theoretical and Mathematical Physics},
  volume = {159},
  number = {1},
  pages = {488--498},
  issn = {1573-9333},
  doi = {10.1007/s11232-009-0039-x},
  language = {en}
}

@article{asano:2014exc,
  title = {Exciton--{{Mott Physics}} in {{Two-Dimensional Electron}}--{{Hole Systems}}: {{Phase Diagram}} and {{Single-Particle Spectra}}},
  shorttitle = {Exciton--{{Mott Physics}} in {{Two-Dimensional Electron}}--{{Hole Systems}}},
  author = {Asano, Kenichi and Yoshioka, Takuya},
  year = 2014,
  journal = {Journal of the Physical Society of Japan},
  volume = {83},
  number = {8},
  pages = {084702},
  publisher = {The Physical Society of Japan},
  issn = {0031-9015},
  doi = {10.7566/JPSJ.83.084702}
}

@article{kumar:2025int,
  title = {Interacting Mesons as Degrees of Freedom in a Chiral Model},
  author = {Kumar, Rajesh and Grefa, Joaquin and Maslov, Konstantin and Wang, Yuhan and Kumar, Arvind and Rapp, Ralf and Ratti, Claudia and Dexheimer, Veronica},
  year = 2025,
  journal = {Physical Review D},
  volume = {111},
  number = {7},
  pages = {074029},
  publisher = {American Physical Society},
  doi = {10.1103/PhysRevD.111.074029}
}

@article{lopes:2022exc,
  title = {Excitonic Insulators and {{Gross-Neveu}} Models},
  author = {Lopes, Nei and Continentino, Mucio A. and Barci, Daniel G.},
  year = 2022,
  journal = {Physical Review B},
  volume = {105},
  number = {16},
  eprint = {2112.07362},
  primaryclass = {cond-mat},
  pages = {165125},
  issn = {2469-9950, 2469-9969},
  doi = {10.1103/PhysRevB.105.165125},
  archiveprefix = {arXiv}
}

@article{steinhoff:2017exc,
  title = {Exciton Fission in Monolayer Transition Metal Dichalcogenide Semiconductors},
  author = {Steinhoff, A. and Florian, M. and R{\"o}sner, M. and Sch{\"o}nhoff, G. and Wehling, T. O. and Jahnke, F.},
  year = 2017,
  journal = {Nature Communications},
  volume = {8},
  number = {1},
  pages = {1166},
  publisher = {Nature Publishing Group},
  issn = {2041-1723},
  doi = {10.1038/s41467-017-01298-6},
  copyright = {2017 The Author(s)},
  language = {en}
}

@article{wrona:2022pai,
  title = {A {{Pair}} of {{2D Quantum Liquids}}: {{Investigating}} the {{Phase Behavior}} of {{Indirect Excitons}}},
  shorttitle = {A {{Pair}} of {{2D Quantum Liquids}}},
  author = {Wrona, Paul R. and Rabani, Eran and Geissler, Phillip L.},
  year = 2022,
  journal = {ACS Nano},
  volume = {16},
  number = {9},
  pages = {15339--15346},
  issn = {1936-0851},
  doi = {10.1021/acsnano.2c06947},
  pmcid = {PMC9527805},
  pmid = {36069715}
}

@misc{Mahato:2025gh,
  author       = {Biplab Mahato},
  title        = {biplab37/PhaseGN: First Release},
  year         = {2025},
  publisher    = {Zenodo},
  version      = {v1.0.0},
  doi          = {10.5281/zenodo.16698543},
  url          = {https://doi.org/10.5281/zenodo.16698543},
}

@article{Levinson:1949os,
  author       = {Levinson, N},
  title        = {On the uniqueness of the potential in a Schrodinger equation for a given asymptotic phase},
  url          = {https://www.osti.gov/biblio/4434712},
  journal      = {Kgl. Danske Videnskab Selskab. Mat. Fys. Medd.  },
  volume       = {Vol: 25, No. 9},
  place        = {Country unknown/Code not available},
  year         = {1949},
  month        = {01}
}

@article{Rosenstein:1988dj,
    author = "Rosenstein, B. and Warr, B. J. and Park, S. H.",
    title = "{Thermodynamics of (2+1)-dimensional Four Fermi Models}",
    reportNumber = "UTTG-27-88",
    doi = "10.1103/PhysRevD.39.3088",
    journal = "Phys. Rev. D",
    volume = "39",
    pages = "3088",
    year = "1989"
}

@article{Rosenstein:1988pt,
    author = "Rosenstein, Baruch and Warr, Brian J. and Park, Seon H.",
    title = "{The Four Fermi Theory Is Renormalizable in (2+1)-Dimensions}",
    reportNumber = "UTTG-22-88",
    doi = "10.1103/PhysRevLett.62.1433",
    journal = "Phys. Rev. Lett.",
    volume = "62",
    pages = "1433--1436",
    year = "1989"
}

@article{Novoselov:2004xxs,
    author = "Novoselov, K. S. and Geim, A. K. and Morozov, S. V. and Jiang, D. and Zhang, Y. and Dubonos, S. V. and Grigorieva, I. V. and Firsov, A. A.",
    title = "{Electric Field Effect in Atomically Thin Carbon Films}",
    doi = "10.1126/science.1102896",
    journal = "Science",
    volume = "306",
    number = "5696",
    pages = "666--669",
    year = "2004"
}

@article{Eguchi:1976iz,
    author = "Eguchi, Tohru",
    title = "{A New Approach to Collective Phenomena in Superconductivity Models}",
    reportNumber = "EFI 76/20-CHICAGO",
    doi = "10.1103/PhysRevD.14.2755",
    journal = "Phys. Rev. D",
    volume = "14",
    pages = "2755",
    year = "1976"
}

@article{Novoselov:2005kj,
    author = "Novoselov, K. S. and Geim, A. K. and Morozov, S. V. and Jiang, D. and Katsnelson, M. I. and Grigorieva, I. V. and Dubonos, S. V. and Firsov, A. A.",
    title = "{Two-dimensional gas of massless Dirac fermions in graphene}",
    eprint = "cond-mat/0509330",
    archivePrefix = "arXiv",
    doi = "10.1038/nature04233",
    journal = "Nature",
    volume = "438",
    pages = "197",
    year = "2005"
}

@article{Kikkawa:1976fe,
    author = "Kikkawa, Keiji",
    title = "{Quantum Corrections in Superconductor Models}",
    reportNumber = "OS-GE 76-1",
    doi = "10.1143/PTP.56.947",
    journal = "Prog. Theor. Phys.",
    volume = "56",
    pages = "947",
    year = "1976"
}

@article{Nambu:1961tp,
    author = "Nambu, Yoichiro and Jona-Lasinio, G.",
    editor = "Eguchi, T.",
    title = "{Dynamical Model of Elementary Particles Based on an Analogy with Superconductivity. 1.}",
    doi = "10.1103/PhysRev.122.345",
    journal = "Phys. Rev.",
    volume = "122",
    pages = "345--358",
    year = "1961"
}

@article{CamaraPereira:2020ipu,
    author = "C{\^a}mara Pereira, Renan and Costa, Pedro",
    title = "{One-meson-loop NJL model: Effect of collective and noncollective excitations on the quark condensate at finite temperature}",
    eprint = "2003.08430",
    archivePrefix = "arXiv",
    primaryClass = "hep-ph",
    doi = "10.1103/PhysRevD.101.054025",
    journal = "Phys. Rev. D",
    volume = "101",
    number = "5",
    pages = "054025",
    year = "2020"
}

@article{Rochev:2009zz,
    author = "Rochev, V. E.",
    title = "{Meson contributions in the Nambu-Jona-Lasinio model}",
    doi = "10.1007/s11232-009-0039-x",
    journal = "Theor. Math. Phys.",
    volume = "159",
    pages = "488--498",
    year = "2009"
}

@article{Zhuang:1994dw,
    author = "Zhuang, P. and Hufner, J. and Klevansky, S. P.",
    title = "{Thermodynamics of a quark - meson plasma in the Nambu-Jona-Lasinio model}",
    doi = "10.1016/0375-9474(94)90743-9",
    journal = "Nucl. Phys. A",
    volume = "576",
    pages = "525--552",
    year = "1994"
}

@article{Maslov:2023boq,
    author = "Maslov, Konstantin and Blaschke, David",
    title = "{Effect of mesonic off-shell correlations in the PNJL equation of state}",
    eprint = "2301.09882",
    archivePrefix = "arXiv",
    primaryClass = "hep-ph",
    doi = "10.1103/PhysRevD.107.094010",
    journal = "Phys. Rev. D",
    volume = "107",
    number = "9",
    pages = "094010",
    year = "2023"
}

@article{Hufner:1994ma,
    author = "Hufner, J. and Klevansky, S. P. and Zhuang, P. and Voss, H.",
    title = "{Thermodynamics of a quark plasma beyond the mean field: A generalized Beth-Uhlenbeck approach}",
    doi = "10.1006/aphy.1994.1080",
    journal = "Annals Phys.",
    volume = "234",
    pages = "225--244",
    year = "1994"
}

@article{Zhukovsky:2015ncz,
    author = "Zhukovsky, V. Ch. and Klimenko, K. G.",
    title = "{The Phase Structure of a Generalized Gross{\textendash}Neveu Model in (2+1)-Dimensional Space{\textendash}Time}",
    doi = "10.3103/S0027134915060181",
    journal = "Moscow Univ. Phys. Bull.",
    volume = "70",
    number = "6",
    pages = "466--472",
    year = "2015"
}

@article{Gusynin:2007ix,
    author = "Gusynin, V. P. and Sharapov, S. G. and Carbotte, J. P.",
    title = "{AC conductivity of graphene: from tight-binding model to 2+1-dimensional quantum electrodynamics}",
    eprint = "0706.3016",
    archivePrefix = "arXiv",
    primaryClass = "cond-mat.mes-hall",
    reportNumber = "NSF-KITP-07-126",
    doi = "10.1142/S0217979207038022",
    journal = "Int. J. Mod. Phys. B",
    volume = "21",
    pages = "4611--4658",
    year = "2007"
}

@article{Gross:1974jv,
    author = "Gross, David J. and Neveu, Andre",
    title = "{Dynamical Symmetry Breaking in Asymptotically Free Field Theories}",
    reportNumber = "COO-2220-19",
    doi = "10.1103/PhysRevD.10.3235",
    journal = "Phys. Rev. D",
    volume = "10",
    pages = "3235",
    year = "1974"
}

@article{Baym:1961zz,
    author = "Baym, Gordon and Kadanoff, Leo P.",
    title = "{Conservation Laws and Correlation Functions}",
    doi = "10.1103/PhysRev.124.287",
    journal = "Phys. Rev.",
    volume = "124",
    pages = "287--299",
    year = "1961"
}

@article{Ebert:2018dzs,
    author = "Ebert, Dietmar and Blaschke, David",
    title = "{Thermodynamics of a generalized graphene-motivated (2+1) D Gross{\textendash}Neveu model beyond the mean field within the Beth{\textendash}Uhlenbeck approach}",
    eprint = "1811.07109",
    archivePrefix = "arXiv",
    primaryClass = "cond-mat.mes-hall",
    doi = "10.1093/ptep/ptz110",
    journal = "PTEP",
    volume = "2019",
    number = "12",
    pages = "123I01",
    year = "2019"
}

@article{Mahato:2024fta,
    author = "Mahato, Biplab and Blaschke, David and Ebert, Dietmar",
    title = "{Beth-Uhlenbeck equation for the thermodynamics of fluctuations in a generalized (2+1)D Gross-Neveu model}",
    eprint = "2409.10507",
    archivePrefix = "arXiv",
    primaryClass = "cond-mat.mes-hall",
    doi = "10.1103/s5jq-tgz6",
    journal = "Phys. Rev. D",
    volume = "112",
    number = "3",
    pages = "036013",
    year = "2025"
}

@article{Blaschke:2025eyt,
    author = {Blaschke, David and R{\"o}pke, Gerd and Baym, Gordon},
    title = "{Generalized Beth--Uhlenbeck entropy formula from the $\Phi-$derivable approach}",
    eprint = "2512.03876",
    archivePrefix = "arXiv",
    primaryClass = "nucl-th",
    month = "12",
    year = "2025"
}

@article{Blaschke:2017boi,
    author = "Blaschke, D. and Ebert, D.",
    title = "{Variational path-integral approach to back-reactions of composite mesons in the Nambu{\textendash}Jona-Lasinio model}",
    eprint = "1703.08964",
    archivePrefix = "arXiv",
    primaryClass = "hep-ph",
    doi = "10.1016/j.nuclphysb.2017.06.013",
    journal = "Nucl. Phys. B",
    volume = "921",
    pages = "753--768",
    year = "2017"
}



%


\PublishersNote{}
\end{document}